\documentclass[preprint,12pt]{elsarticle}
\usepackage{lineno,hyperref}
\usepackage{xcolor}

\usepackage{microtype} 
\usepackage[all]{nowidow} 
\usepackage{siunitx} 
\usepackage{textcomp} 
\journal{Journal of Heat and Mass Transfer}
\graphicspath{{./images/}}
\begin{document}
\begin{frontmatter}
\title{Model geometries of porous materials}
\author{Felix Buchele\corref{cor1}}
\ead{felix.buchele@fau.de}
\author{Patric Müller}
\ead{patric.mueller@fau.de}
\author{Michael Blank}
\ead{michael.u.blank@fau.de}
\author{Thorsten Pöschel}
\ead{thorsten.poeschel@fau.de}
\cortext[cor1]{Corresponding author}
\address{Friedrich-Alexander-Universität Erlangen-Nürnberg, Institute for Multiscale Simulation (MSS),
        Cauerstraße 3,
        91058 Erlangen,
        Germany}
\begin{abstract}
We describe a method for modeling the geometry of porous materials. The approach enables the independent selection of crucial parameters, including porosity, pore size distribution, pore shape, and connectivity. Consequently, it can effectively model a wide range of porous systems. Due to the diverse and systematic variation possibilities, the method is suitable for developing and optimizing porous structures. The geometries can be exported as triangular meshes, facilitating their immediate use in numerical simulation and further digital processing.  
We showcase the method's capabilities by minimizing the foam structure's thermal conductivity through geometry optimization.
\end{abstract}



\end{frontmatter}
\newpage
\setcounter{page}{1}
\section{Introduction}
Porous materials play a role in many scientifically interesting and technically relevant systems. Depending on the question, porosity can be considered on different levels of accuracy. For some questions in the field of geosciences, for example, effective descriptions in terms of macroscopic laws, such as e.g. Darcy's law, are sufficient. Other highly topical issues related to heat and mass transfer in porous media, such as catalysis, thermal insulation, and filtering, require understanding at the pore-scale level. In this case, the geometry of the porous structures must be considered explicitly. The system geometries can be obtained experimentally by 3d scanning techniques like computed tomography. However, for the development and optimization of porous structures using numerical simulation, versatile digital model geometries are needed that, on the one hand, can accurately describe realistic systems and, on the other hand, can be systematically varied and parameterized \cite{Meakin.2009, Xiong.2016}.

\begin{figure}[!b]
    \centering
    \includegraphics[width=\textwidth]{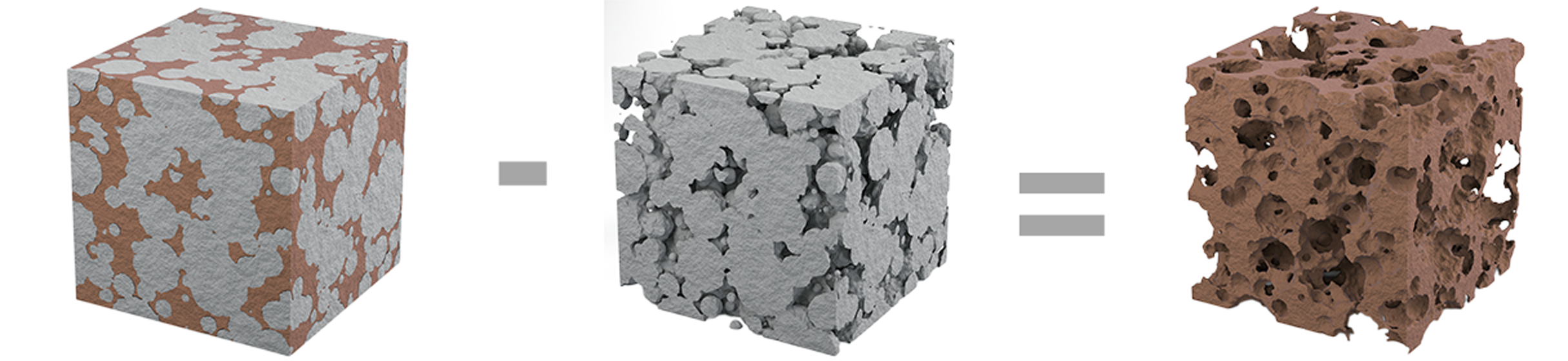}
    \caption{Sphere arrangements and inverse sphere arrangements as models for solid phase and void phase porous materials.}
    \label{fig:invpack}
\end{figure}

Various methods for generating porous geometries on different levels of simplification are described in the literature. Monte Carlo methods, for example, where pores are subtracted from a solid geometry until a desired porosity is reached, have been employed by Torquato \cite{Torquato.1991} and De Cariolis \cite{DeCarolis.2023}. Other simplifications include the replacement of pores by simpler geometries such as interconnected tubes \cite{Liu.2022}, methods, where pores are placed randomly and grow until desired pore metrics are reached \cite{Wang.2007}, or purely statistical methods such as multiple-point statistics \cite{Wang.2018}. Representative elementary volumes (REV's) are frequently employed for large-scale simulations.
A more sophisticated representation is required for other problems, such as the reactive flow on chemically active surfaces of a porous catalyst. An example is given in \cite{Muhlbauer.2019}, where the pore space of a catalyst is represented by repeated unit cells. Dyck and Straatman \cite{Dyck.2015} obtain porous geometries from random close packings of spheres, which, in turn, have been obtained from DEM simulations. An overview of various methods to create representations for porous systems is given, for example, by Xiong et al. \cite{Xiong.2016}, Al-Raoush et al \cite{AlRaoush.2003}, Golparvar et al \cite{Golparvar.2018} and Chen et al \cite{Chen.2022}.

The models available in the literature are generally tailored to \emph{specific} applications, which they then describe in good quality. However, none of the existing methods meet the criteria relevant to the numerical optimization of porous structures, i.e., they are versatile, realistic, and can be systematically varied by parameters. The method presented in this paper is intended to fill this gap. To that, we extend the approach by Dyck and Straatmann \cite{Dyck.2015} to aspherical pores with arbitrary overlaps and post-processing of the individual pores by, e.g., texturing the pores' surface.

We showcase our method by a) replicating a natural porous material, simulating its thermal conductivity, and validating it with measured data and by b) minimizing the heat conductivity of a fictional open-porous material with overlapping, non-spherical pores at constant porosity. Such porous structures appear, for example, in macro-porous ceramics, which can be produced by directly foaming a ceramic suspension or by using a precursor with additives that lead to the formation of pores through pyrolysis during firing.

\section{Modified sphere packings as a model for porous geometries\label{sec:method}}
\begin{figure}
    \centering
    \includegraphics[width=\textwidth]{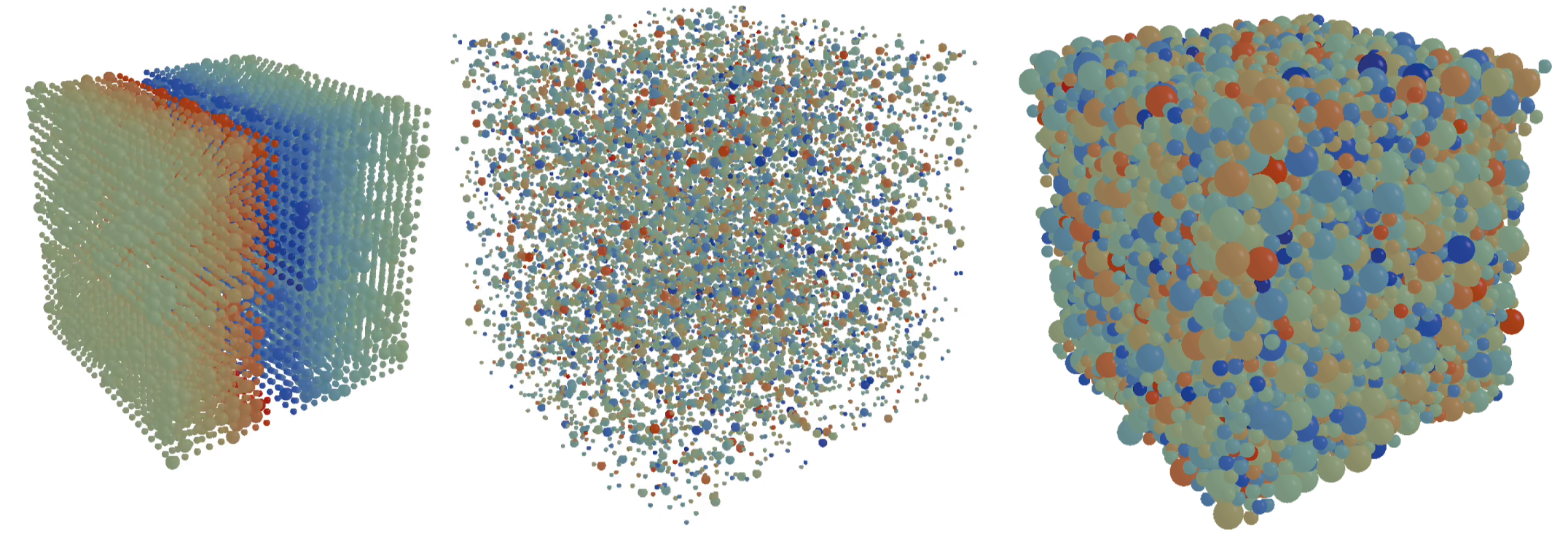}
    \caption{Positioning of the geometric primitives (here spheres) in space. \emph{Left:} Initial position of the spheres on a rectangular lattice in a cubic domain. \emph{Middle:} Position after equilibration phase. \emph{Right:} Result of the Lubachevsky-Stillinger compression.}
    \label{fig:LS-algorithm}
\end{figure}

One way to model porous matter is to compose the solid matrix by placing elementary geometric bodies like spheres in space. This arrangement can be used directly as a model for a solid-phase porous material. Furthermore, the union of all bodies, following constructive solid geometry principles, can be subtracted from an encompassing solid body to obtain a void phase porous geometry as shown in Fig. \ref{fig:invpack}. In both cases, the elementary geometric bodies can be modified in various ways as an intermediate step. Such modifications include scaling, deformation, changing the position and orientation, or adding a surface texture. Our method for generating models of porous structures, therefore, decomposes into three main steps:

\begin{figure}
    \centering
    \includegraphics[width=\textwidth]{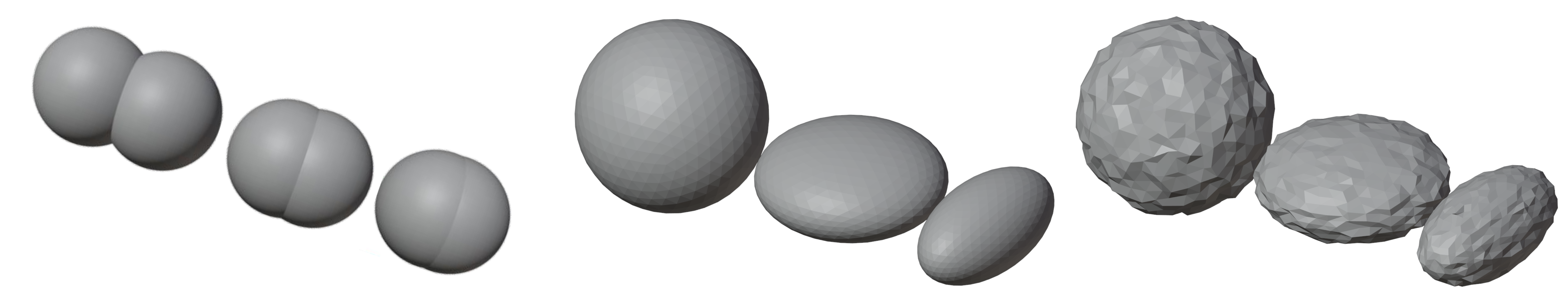}
    \caption{Modification of the geometric primitives in size (left), shape (middle), and surface texture (right).\label{fig:modPrimitves}}
\end{figure}

\begin{figure}
    \centering
    \includegraphics[width=0.8\textwidth]{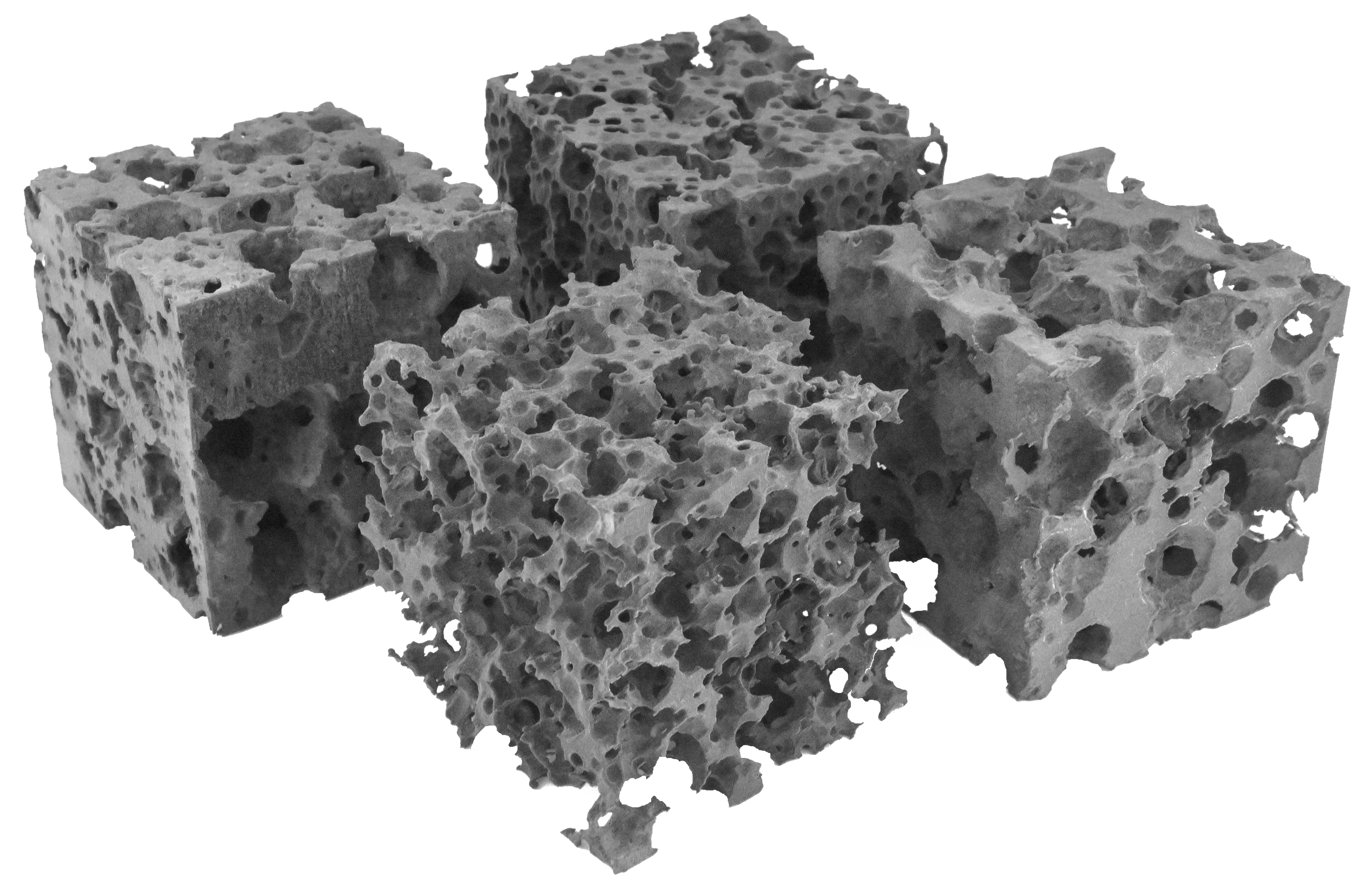}
    \caption{Exemplary open porous geometries resulting from the described modified sphere packing algorithm.\label{fig:exampleGeometries}}
\end{figure}

\begin{enumerate}
    \item \emph{Positioning of the geometric primitives in space:}
    Typically, a large number of objects need to be distributed here, so automated methods are usually required. In this work, we restrict ourselves to spherical objects and choose event-driven DEM (e.g., \cite{Bannerman.2011}) for the automatic random positioning. As shown later, the random spatial distribution of the pores is a good approximation of many scientifically interesting or technically relevant systems. First, we set up a cuboidal simulation domain with periodic boundary conditions containing the desired number of spheres. The radii of the spheres are chosen according to a desired pore size distribution. In the first simulation step, we equilibrate the system by specifying random velocities for the particles, assuming elastic interactions, and then simulating their dynamics for a defined period (e.g., 50 collisions per particle). Next, we adjust the packing fraction by applying the Lubachevsky-Stillinger algorithm \cite{Lubachevsky.1990}, where the diameter of each sphere is increased at a specified rate until the desired packing fraction, i.e., porosity, is reached. Figure \ref{fig:LS-algorithm} shows this three-stage process.
    \item \emph{Modification of the geometric primitives (if required):} To carry out various modifications, we represent each geometric primitive by individual triangular meshes from here on. Fig. \ref{fig:modPrimitves} exemplary shows the variation of object size, shape, and surface texture. By scaling the objects, their overlap can be tuned. This makes it possible to choose between open- and closed-porous systems. In this work, we use the meshing tool \textit{pymesh} \cite{Zhou.2018} to perform the mesh modifications. 
    \item \emph{Generation of the inverse packing (in case of void phase porous geometries):}
    The result of the previous step is solid-phase porous structures. Suppose we subtract the sum of all individual geometric primitives from a piece of solid material in the sense of constructive solid geometry. In that case, we arrive at the complementary void-phase porous structure. 
\end{enumerate}
The described procedure makes it possible to create a wide variety of porous structures. Fig. \ref{fig:exampleGeometries} shows some examples of solid and complementary void phase porous geometries. Further examples are discussed later in the application demonstrations in Sec. \ref{sec:demoA} and Sec. \ref{sec:demoB}.

\section{Application demonstration A: Replication of a ceramic foam structure\label{sec:demoA}}
This Section demonstrates the modified sphere packing method by replicating the ceramic foam structure. To validate the result, we simulate heat transport through the artificial porous geometry and compare the resulting heat conductivity to that of natural ceramic foam obtained from experiments.

\begin{figure}[!b]
    \centering
    \includegraphics[width=\textwidth]{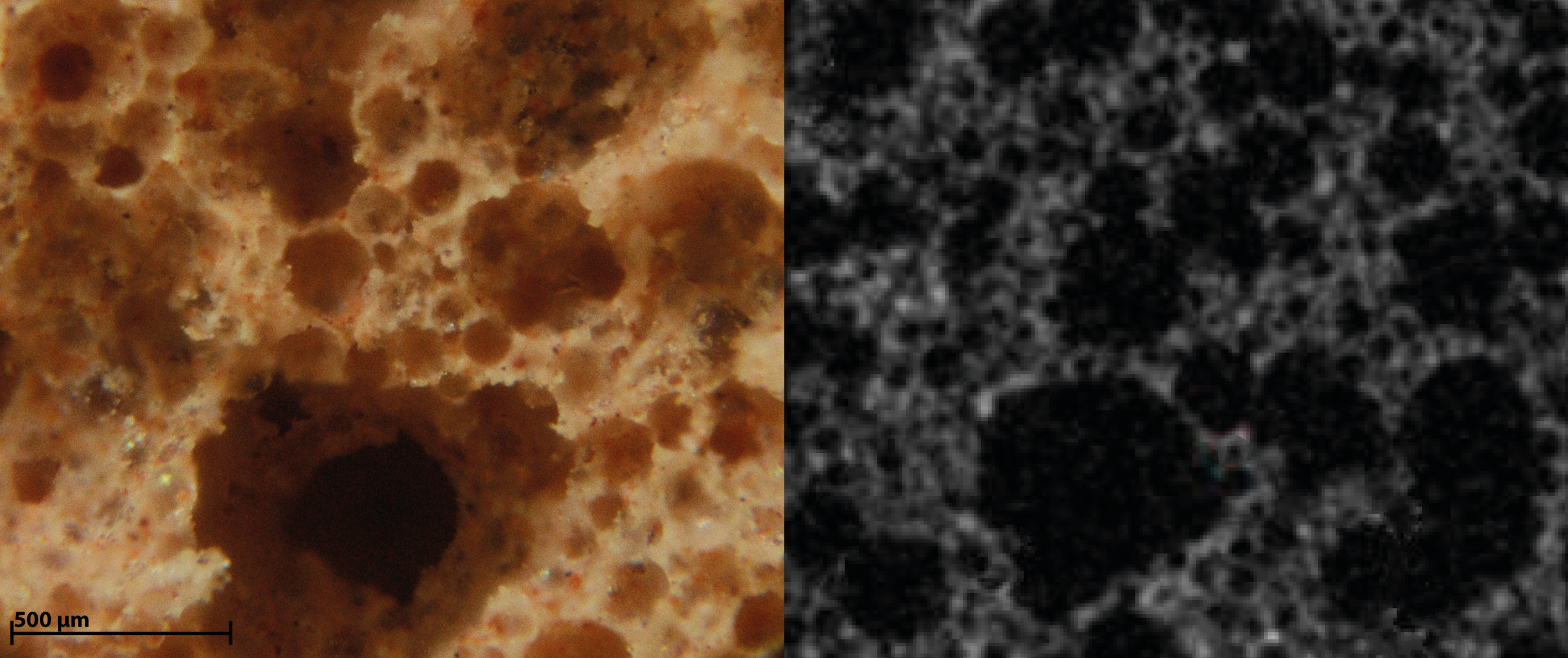}
    \caption{Microscopy image of a porous ceramic (left) and CT image data obtained from the same sample}
    \label{fig:microscopy-CT}
\end{figure}

\begin{figure}[!t]
    \centering
    \includegraphics[width=\textwidth]{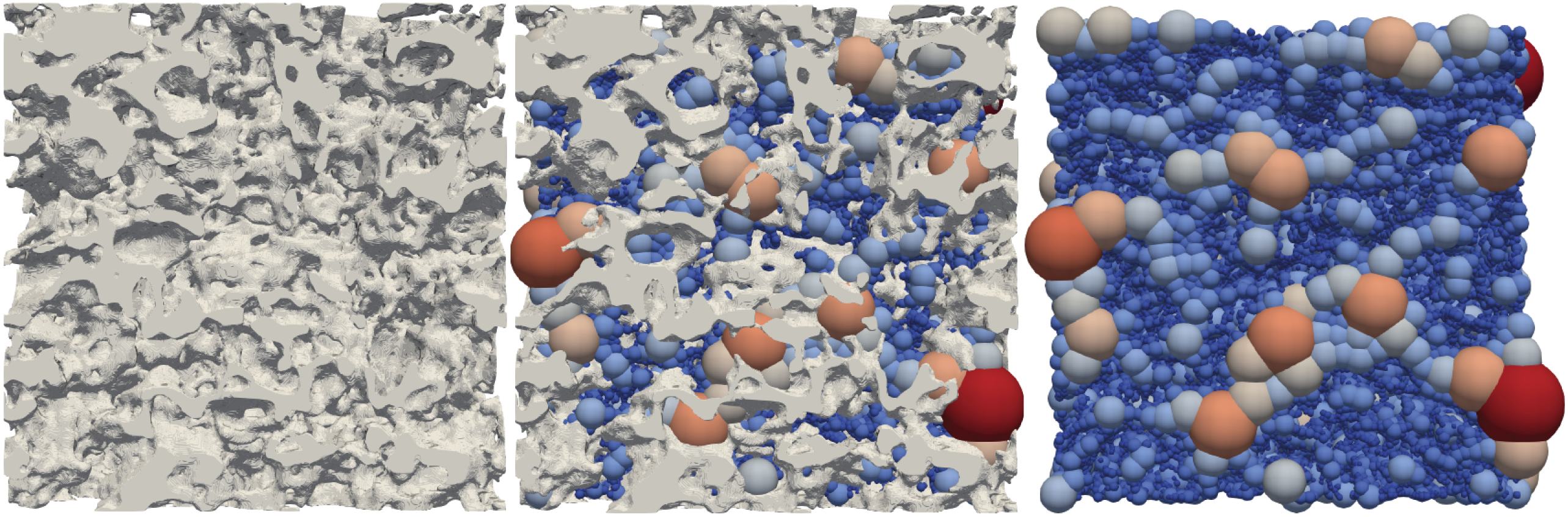}
    \caption{$\mu$-CT-data (left), pores detected according to the method described in \cite{Angelidakis.2021} (right), and overlay of both (center)}
    \label{fig:sphere_fitting}
\end{figure}

\begin{figure}[!b]
    \centering
    \includegraphics[width=0.8\textwidth]{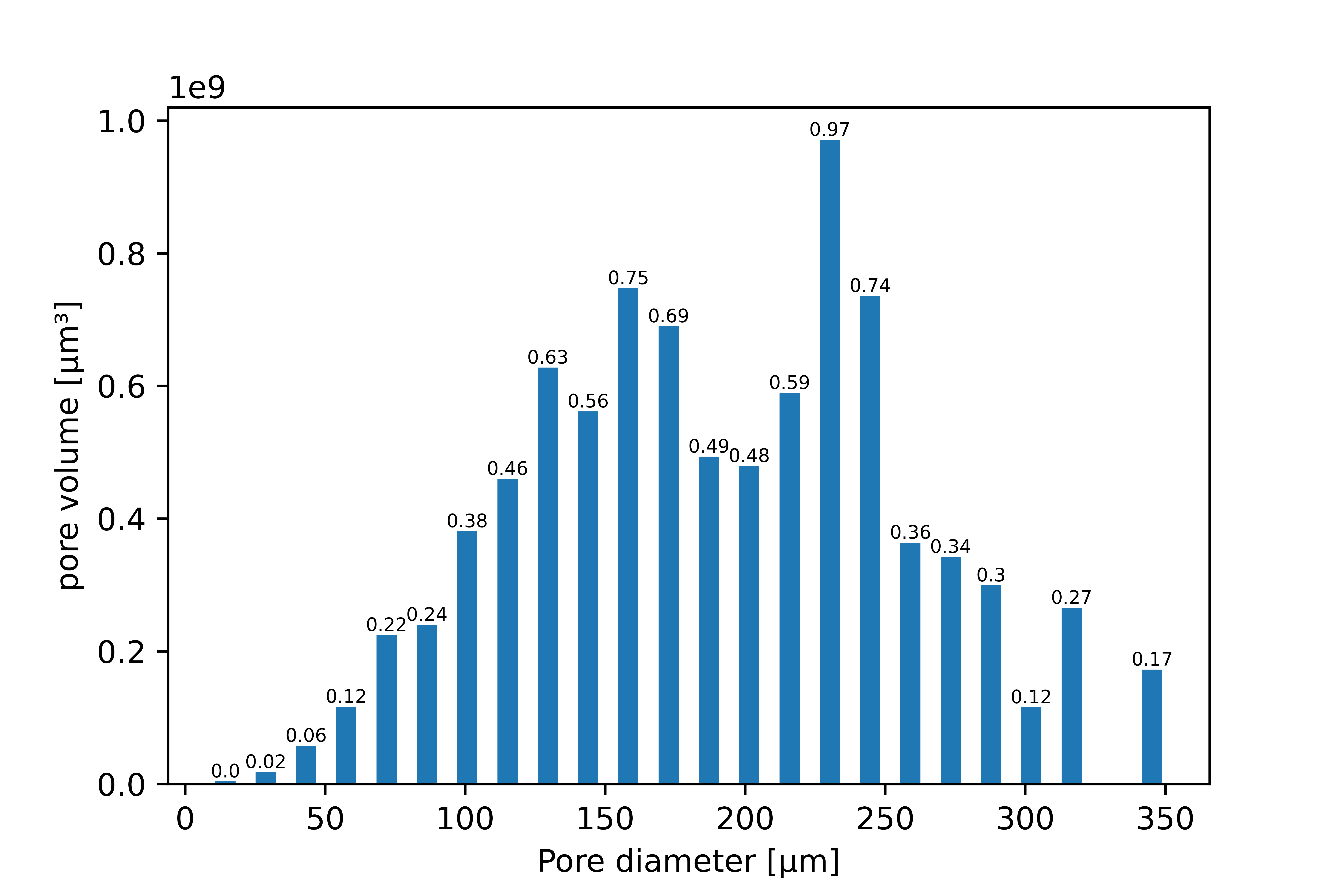}
    \caption{Pore volume per diameter as a function of the pore diameter.}
    \label{fig:pore-metrics}
\end{figure}

To replicate the foam structure using the method described in Sec. \ref{sec:method}, the pore size distribution, the porosity, the overlap ratio between neighboring pores, and the sphericity of the pores are required. We perform a $\mu$-CT analysis of a \SI{1}{\cubic\centi\meter} sample of the ceramic foam to obtain these metrics. Fig. \ref{fig:microscopy-CT} shows microscopy images of the sample and a 2D slice of the acquired tomogram.
As we can see from both the microscopy data and the CT data, all pores are almost perfectly spherical, and the pores' positions and sizes are distributed randomly over space. Additionally, pores frequently combine and overlap. All these observations result from the foaming process where randomly placed gas cavities emerge during firing due to additives randomly distributed over the raw material. Therefore, we can safely assume a pore sphericity of 1 and use the method described in \cite{Angelidakis.2021} to obtain all further metrics of the porous structure required to replicate the geometry. This method was initially intended to represent complicated shapes by assemblies of spheres. Still, it can easily be modified to fit spheres to the void space of a porous structure and, therefore, obtain the position and size of the spherical pores. Figure \ref{fig:sphere_fitting} shows the result of this procedure.
From the size and the positions of the detected pores, we can directly obtain the pore size distribution, the overlap between neighboring pores, and the porosity. Alternatively, the porosity can be obtained by binarizing the 3D-CT data and computing the void ratio.
Figure \ref{fig:pore-metrics} shows the measured pore size distribution and the average overlap between pores as a function of the pore diameter. The porosity is approximately \SI{60}{\percent}. Remarkably, the pore sizes follow an exponential distribution.

\begin{figure}[!t]
    \centering
    \includegraphics[width=\textwidth]{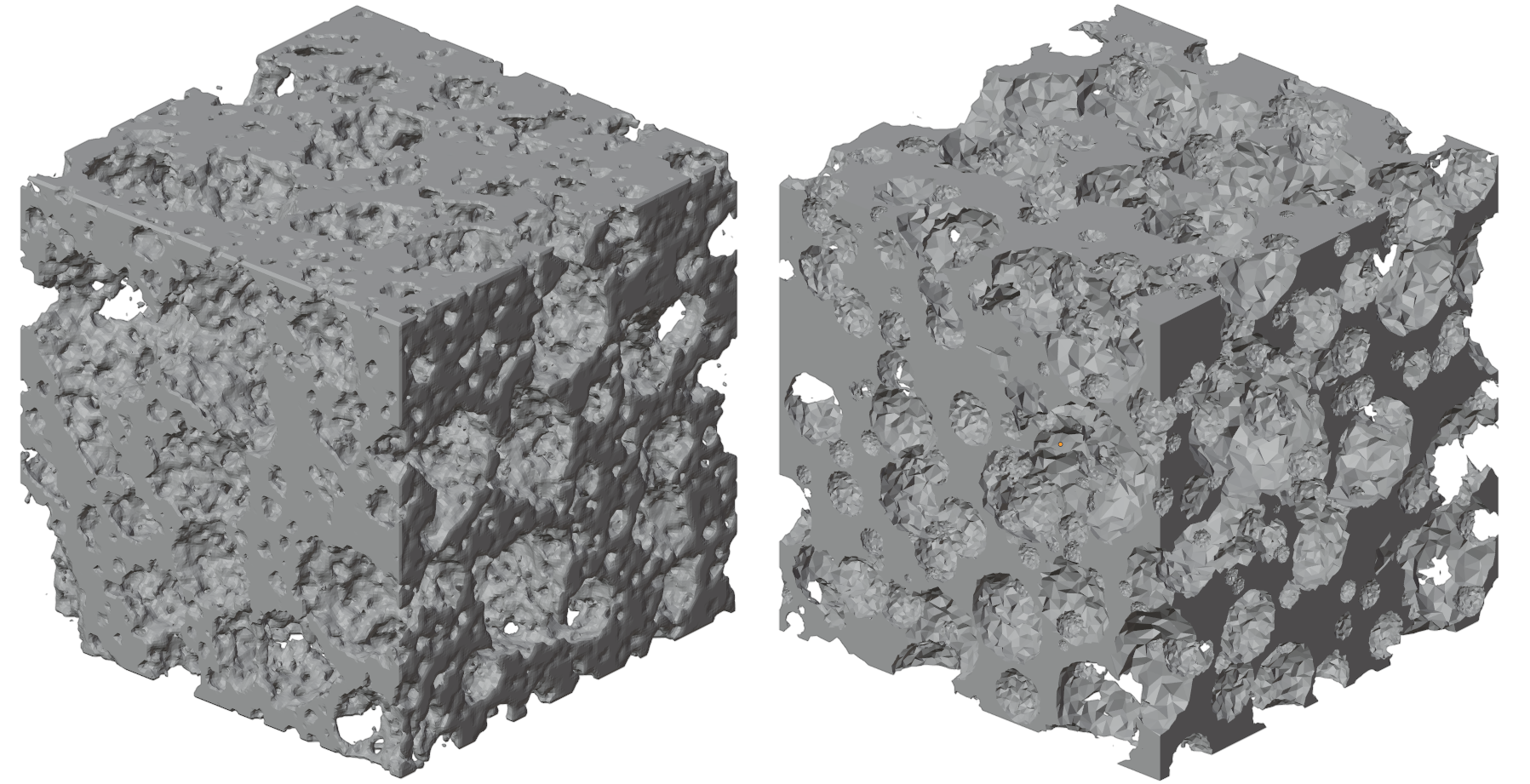}
    \caption{Geometry of a ceramic foam obtained by $\mu$-CT (left). Replication of the geometric properties using the method described in Sec. \ref{sec:method} (right)}
    \label{fig:tomogram-replication}
\end{figure}

Using the obtained characteristics of the ceramic foam, we now apply the method described in Sec. \ref{sec:method} to create an artificial porous structure with the same statistical properties as the natural material. While we can precisely replicate the pore size distribution, the porosity and overlap between pores can only be replicated within a tolerance of \textpm \SI{3}{\percent} and \textpm \SI{5}{\percent}, respectively. A qualitative comparison between the natural material and the replication can be seen in Figure \ref{fig:tomogram-replication}. The geometries there consist of more than 100.000 individual pores and represent a cubic cutout of the ceramic foam with a side length of \SI{200}{\micro\metre}.

To further validate our method for reproducing porous geometries, we conduct heat transfer experiments on a sample material and compare the results with heat transfer simulations of
the replicated geometry.

In the experiment, we measure a ceramic foam panel with the dimensions  $200 \times 200 \times 20$ \si{\milli\metre}. To obtain the heat conductivity, the sample is clamped between two plates. One of the plates is kept at a temperature of \SI{15}{\celsius}, and the other at \SI{5}{\celsius}. By measuring the heat flux  for the given temperature gradient, the heat conductivity can be obtained directly from the Fourier law:
\begin{equation}
\lambda=\frac{\dot{Q}}{A}\frac{d}{T_\text{hot}-T_\text{cold}}
\end{equation}
where $\dot{Q}$ is the heat output of the hot plate, $A$ is the surface area of the plate, $d$ is the thickness of the sample, and $T_\text{hot}$ and $T_\text{cold}$ are the temperature of the hot and the cold plate respectively. The sides of the sample are thermally insulated such that there is no heat flux directed parallel to the surface of the hot and cold plate. From this measurement, we obtain the thermal conductivity \SI{56.9}{\milli\watt\per\metre\per\kelvin}.

\begin{figure}[htbp]
    \centering
    \includegraphics[width=0.55\textwidth]{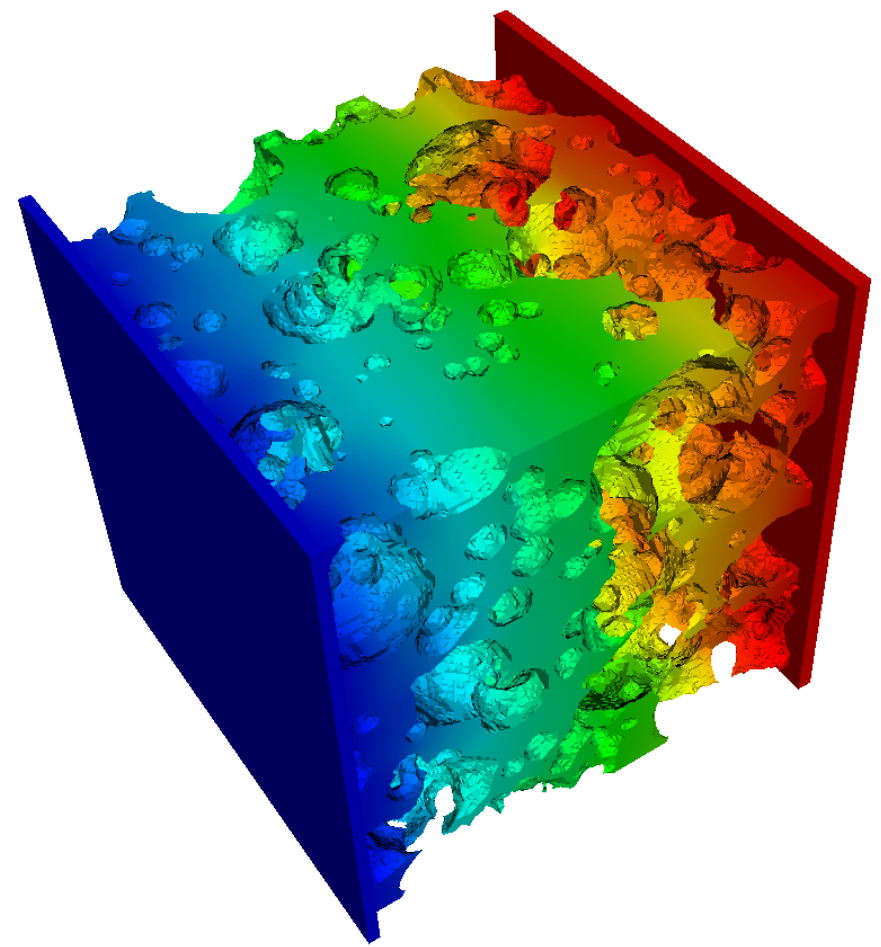}
    \caption{Setup for the heat transfer simulations. The sample is clamped between a hot and cold plate (left and right). At all other faces, we specify Neumann boundary conditions so that no heat transfer occurs. This corresponds to a perfectly isolated system.}
    \label{fig:sph_simulation}
\end{figure}

Four heat conduction mechanisms contribute to the measured overall heat transfer: Heat conduction through the solid matrix, heat conduction through the enclosed fluid, convective heat transport, and heat transfer by radiation. In principle, all four mechanisms must, therefore, also be modeled in a corresponding simulation. However, for small pores, viscous effects dominate and inhibit natural convection. There is an ongoing debate on the critical pore size above which natural convection is significant. In \cite{burnsAndTien}, it was shown that convection does not occur in porous media if the Rayleigh number is smaller than 75. Later studies suggested that the contribution of convection to the effective thermal conductivity is negligible for pore sizes below \SI{4}{\milli\metre} \cite{burnsAndTien,12,13,14,15,16}. Others determined the critical pores size to \SI{3}{\milli\metre} \cite{17} or \SI{5}{\milli\metre} \cite{18}. A nice overview is given in \cite{ZiHao}. As we can see from the measured pore size distribution in Fig. \ref{fig:pore-metrics}, the pore sizes of the ceramic foam are below $\approx$ \SI{400}{\micro\metre} and, thus, definitely far below all critical pore sizes suggested so far. Therefore, we can safely neglect the contribution of convective transport to the effective thermal conductivity of the studied ceramic foam. Hence, in our simulations, we only need to consider heat conduction through the solid and the gas and heat radiation. The heat conduction simulations are done with an in-house implementation of the Smoothed Particle Hydrodynamics method (SPH), see e.g. \cite{BLANK2019}. SPH is a mesh-free Lagrangian method originally introduced for treating astrophysical phenomena and gas dynamics \cite{SPHA,SPHB}. The mesh-free approach has proven to be advantageous for geometrically complicated boundary conditions, as is the case with porous foam structures. The heat radiation is modeled by ray tracing. Each surface element emits thermal energy according to the Stefan-Boltzmann law
\begin{equation}
P=\varepsilon(T)\sigma A T^4
\end{equation}
where $P$ is the radiation power, $\varepsilon$ is the emissivity of the material, $A$ the area of the surface element and $T$ its temperature. We chose the emissivity of brick, $\varepsilon=0.9$ \cite{brewster}. We then assume that the heat energy is radiated along a straight line. When this heat ray hits another surface element, it is absorbed, and its heat energy is transferred to the respective surface element. The reflectivity of the ceramic foam is negligible. For the thermal conductivity of the interstitial air, we use the literature value of \SI{26,2}{\milli\watt\per\metre\per\kelvin} \cite{haynes2014crc}. For the solid matrix, we use the heat conductivity of the unfoamed raw material \SI{395.2}{\milli\watt\per\metre\per\kelvin} as obtained from the experiment. The simulation setup is depicted in Figure \ref{fig:sph_simulation}. Analogous to the experimental setup, the simulation features a cold plate (left) and a hot plate (right) with the same temperatures as in the experiment. Different from the experiment, we simulate 10  cubical samples of the dimension $20 \times 20 \times 20$ \si{\milli\metre} and find an effective thermal conductivity of $56.0\pm0.1$\si{\milli\watt\per\metre\per\kelvin}. The discrepancy to the experimental reference value is approximately \SI{1.5}{\percent}. This consistency showcases the geometric fidelity of the presented method for generating porous structures.

\section{Application demonstration B: Optimization of a porous heat-insulation material\label{sec:demoB}}
\begin{figure}[!ht]
    \centering
    \includegraphics[width=0.7\textwidth]{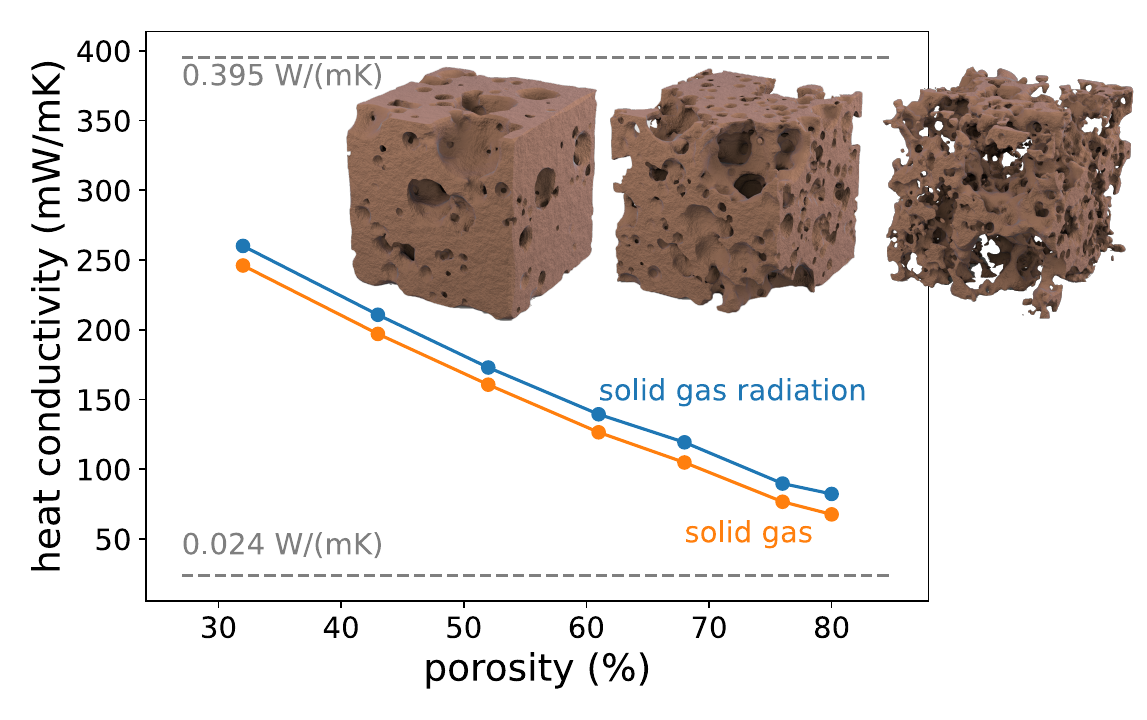}
    \caption{Effective heat conductivity of the foam structure as a function of the porosity. Orange line: heat transport through the solid matrix, interstitial gas, and radiation. Blue line: no radiative transport. The dashed grey lines indicate the air's heat conductivity and the solid matrix's unfoamed raw material. The inset shows the porous geometry for a porosity of \SI{30}{\percent}, \SI{50}{\percent} and \SI{85}{\percent}.
    \label{fig:var-porosity}}
\end{figure}

\begin{figure}[htbp]
    \centering
    \includegraphics[width=0.7\textwidth]{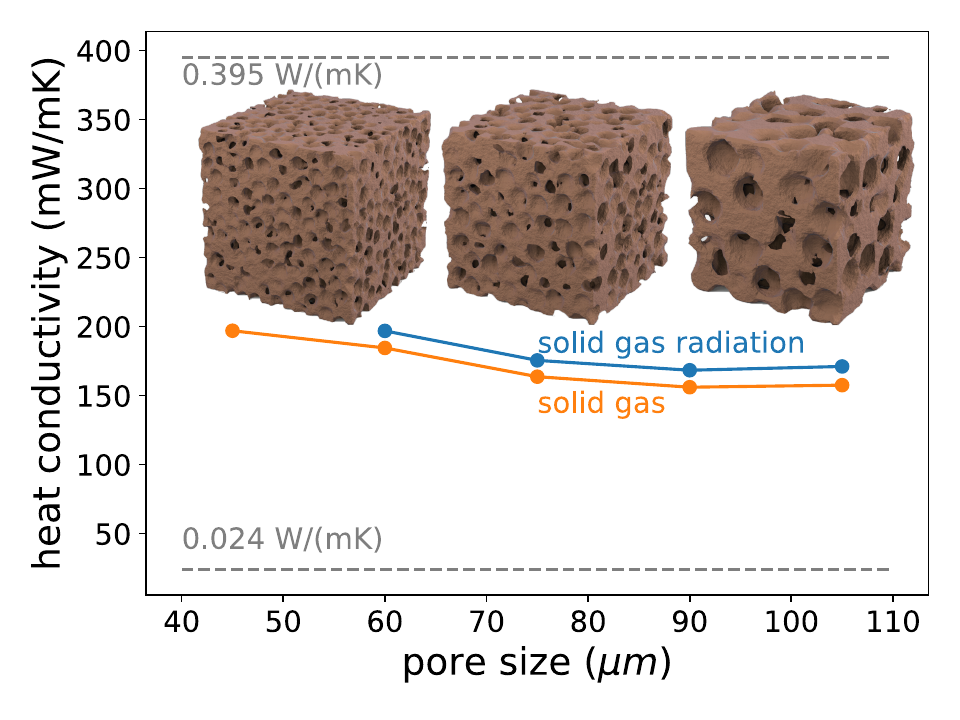}
    \caption{Effective heat conductivity of the foam structure as a function of the average pore size (for further description see caption of Fig. \ref{fig:var-porosity}). The inset shows the porous geometry for an average pore size of \SI{45}{\micro\metre}, \SI{75}{\micro\metre} and \SI{0}{\micro\metre}.
    \label{fig:var-pore-size}}
\end{figure}

In this Section, we demonstrate the variety of possible geometries and show how the presented method for creating porous geometries can be used to vary the produced geometries systematically. To do this, we minimize the thermal conductivity of a fictitious porous thermal insulation material by varying its geometry. We determine the resulting thermal conductivity in each case by simulation as described in Sec. \ref{sec:demoA}. As shown in Fig \ref{fig:var-porosity} and Fig. \ref{fig:var-pore-size}, we can use our method to vary the porosity of the geometry for a given pore size distribution or to vary the average pore size of a porous structure at constant porosity. From numerical simulations described in Sec. \ref{sec:demoA}, we obtain that the heat conductivity decreases almost linearly with increasing porosity. This is to be expected because a higher porosity increases the proportion of air in the material, which in turn has a lower thermal conductivity than the material from which the solid matrix is made. By scaling up the entire porous geometry and cropping the result to the dimensions of the original sample, we can increase the pore size while keeping the porosity constant. Some sample geometries are depicted in the inset of Fig. \ref{fig:var-pore-size}. The resulting heat conductivity decreases with increasing pore size. This is because larger pores prevent or lengthen thermal bridges from the warm to the cold side of the sample more than small pores. However, this is only valid as long as the pore size stays below the threshold at which convection influences the effective heat conductivity of the system (see discussion in Sec. \ref{sec:demoA}).

In practical application, both parameters, the pore size distribution, and the insulator's porosity are frequently predetermined by the manufacturing process and requirements for the mechanical stability of the resulting material. In the following, we, therefore, limit ourselves to the pore size distribution shown in Fig. \ref{fig:pore-metrics} 
\begin{figure}[htbp]
    \centering
    \includegraphics[width=0.7\textwidth,bb=15 15 450 332,clip]{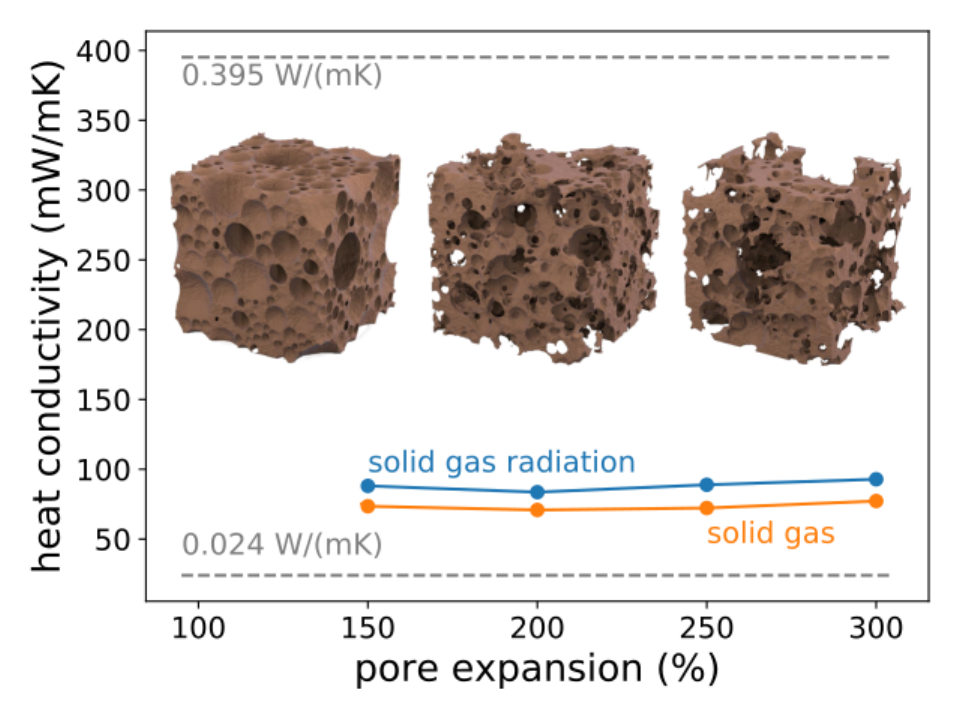}
    \caption{Effective heat conductivity of the foam structure as a function of the pore expansion factor (see text). For further description of the figure, see the caption of Fig. \ref{fig:var-porosity}. The inset shows the porous geometry for a pore expansion of \SI{100}{\percent} (closed porous system), \SI{200}{\percent}, and \SI{300}{\percent}.
    \label{fig:var-open-close}}
\end{figure}
and porosity of \SI{75}{\percent}. Possible degrees of freedom that remain for the optimization of the material include the shape of the individual pores, the orientation of the pores in the case of aspherical pores, or the transition from a closed porous system to an open porous system. To vary from closed to open porous structures at constant porosity, we create closed-cell systems whose porosity is below the target porosity and then enlarge the pores by a given pore expansion factor. This allows us to increase the porosity up to the target value, resulting in overlapping pores, i.e., an open-cell structure. Fig. \ref{fig:var-open-close} shows three exemplary geometries with different pore expansions and the dependence of the effective thermal conductivity on the open cellularity. As we can see from the numerical simulations, the pore overlap hardly influences thermal conductivity. This is because although the pores now overlap, they still do not form continuous channels longer than the critical length scale above, for which convection significantly contributes to heat transport.
\begin{figure}[htbp]
    \centering
    \includegraphics[width=0.7\textwidth,bb=15 15 450 332,clip]{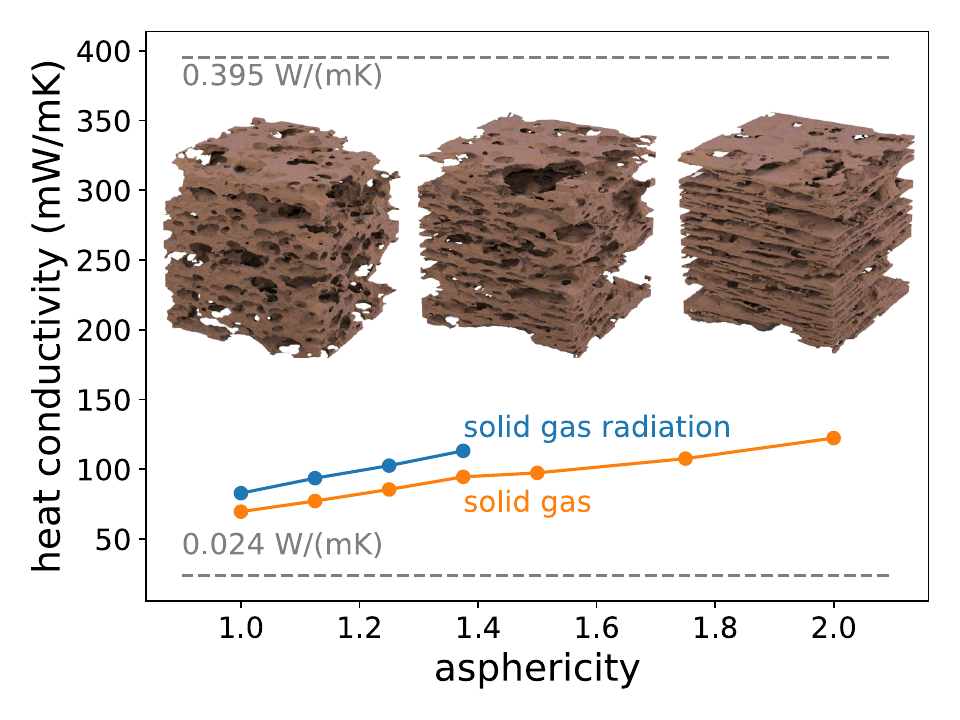}
    \caption{Effective heat conductivity of the foam structure as a function of the pore asphericity. For further description of the figure, see the caption of Fig. \ref{fig:var-porosity}. The inset shows the porous geometry for a pore asphericity of 1.25, 1.5, and 2.0.
    \label{fig:var-sphericity}}
\end{figure}
\begin{figure}[htbp]
    \centering
    \includegraphics[width=0.7\textwidth,bb=80 260 515 625,clip]{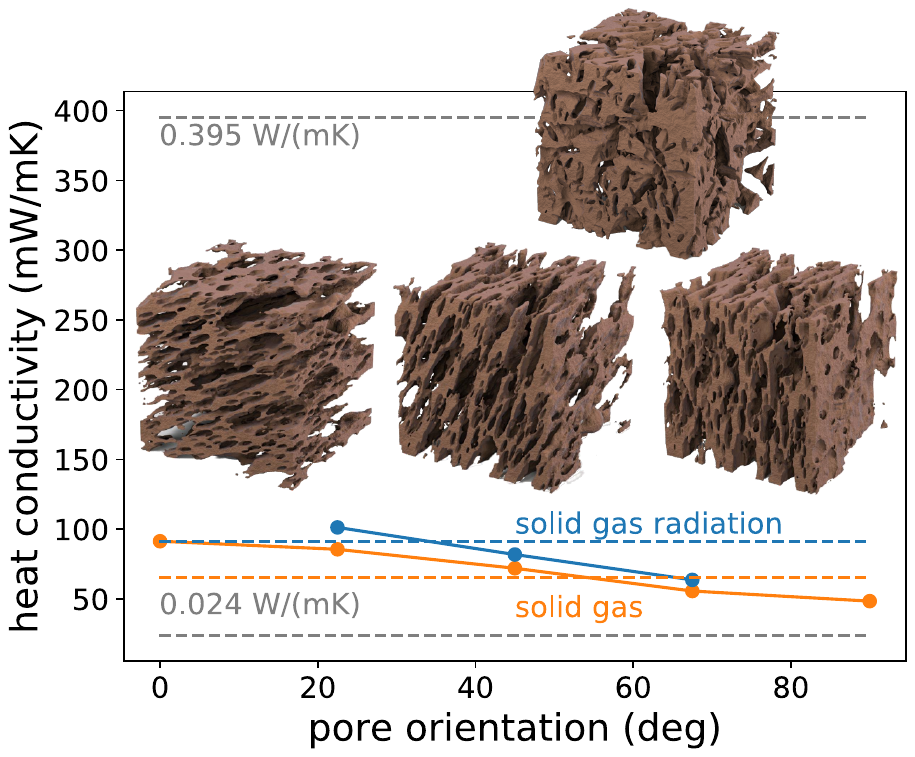}
    \caption{Effective heat conductivity of the foam structure as a function of the orientation of elongated pores relative to the direction of heat transfer. For further description of the figure, see the caption of Fig. \ref{fig:var-porosity}. The inset shows the porous geometry for relative orientations of  \SI{22.5}{\degree}, \SI{45}{\degree} and \SI{90}{\degree}. The image at the top of the inset shows a random orientation of the pores. The dashed orange and blue lines indicate the sample's thermal conductivity with random orientation of the pores.
    \label{fig:var-orientation}}
\end{figure}

To investigate the influence of spherical pores at constant porosity, we consider the initially spherical pores as ellipsoids, reducing one of their semi-axes by a factor and increasing another by the same factor. This way, we obtain elongated or flattened ellipsoidal pores with unchanged volume. Due to the aspherical pores, the porous geometry is no longer isotropic, and we have to determine in which direction we investigate the heat transfer. First, we consider the case where all pores elongate in the same direction. We vary the elongation and look at the heat transport along the longest semi-axis of the ellipsoids. The result is shown in Fig. \ref{fig:var-sphericity}. We can see that the effective heat conductivity increases linearly with the asphericity. This is due to the increasingly layered structure 
where more and more continuous webs form from cold to warm, creating thermal bridges.

We now vary the difference between the direction of the elongated pores and the heat flow direction for an asphericity of 1.5. The more the webs of the layered structure are oriented perpendicular to the direction of the heat flow, the lower the thermal conductivity is, as this avoids thermal bridges. Additionally, we simulated the heat conductivity for a sample with random orientation of the individual pores, as shown in the top row of the inset in Fig. \ref{fig:var-orientation}. This results in an approximate averaging of the thermal conductivities measured for the individual orientations.

From these parameter studies, we can conclude that one way to minimize the thermal conductivity of the porous foam structure at a given porosity and pore size distribution is to flatten the pores and orient the short half-axis of the thus aspherical pores perpendicular to the direction of heat flow.

\section{Summary}
We have introduced a method that allows for generating diverse porous geometries. The technique allows for specifying essential parameters such as porosity, pore shape and orientation, and pore size distribution. In contrast to previously published methods, the parameters can be systematically varied, and highly porous, highly irregular structures with arbitrarily large pore overlap can also be generated. The method is thus suitable for replicating existing natural porous materials and, on the other hand, for conducting parameter studies with fictitious porous structures, as required for the numerical optimization and development of novel porous materials using computer simulations. We validated the method by replicating a ceramic foam structure and comparing the simulated thermal conductivity of the resulting geometry with experimental results. By optimizing the thermal conductivity of a fictitious porous thermal insulation material, we have demonstrated the wide variety of representable geometries and how these can be systematically varied.
\clearpage

\bibliography{Paper_PorousMediaGeneration.bib}
\end{document}